\documentclass[aps,prl,twocolumn,groupedaddress]{revtex4}

\usepackage[latin1]{inputenc}
\usepackage[T1]{fontenc}
\usepackage[english]{babel}
\usepackage[dvips]{graphicx}
\usepackage{amsmath}
\usepackage{enumerate,amsthm,amsmath,amssymb,color,graphicx,bbm,float}
\usepackage{natbib}

\usepackage{epstopdf}

\begin{document}

\title{
Unconditional security proof of long-distance \\
continuous-variable quantum key distribution with discrete modulation.}

\author{Anthony Leverrier}
\affiliation{Institut Telecom / Telecom ParisTech, CNRS LTCI,\\
  46, rue Barrault, 75634 Paris Cedex 13, France} 
\author{Philippe Grangier}
\affiliation{Laboratoire Charles Fabry, Institut d'Optique, CNRS, Univ. Paris-Sud,\\
  Campus Polytechnique, RD 128, 91127 Palaiseau Cedex, France}

\date{\today}

\begin{abstract}
We present a continuous-variable quantum key distribution protocol combining a discrete modulation and reverse reconciliation. This protocol is proven unconditionally secure and allows the distribution of secret keys over long distances, thanks to a reverse reconciliation scheme efficient at very low signal-to-noise ratio. 
\end{abstract}


\maketitle

The first practical application of Quantum Information Theory is
certainly quantum key distribution (QKD) \cite{scarani08}, 
which allows two distant parties to communicate with absolute privacy, even
in the presence of an eavesdropper. Two families of QKD protocols
coexist today, relying either on photon counting techniques or
homodyne detection, which correspond to discrete and
continuous-variable protocols, respectively. The better efficiency of homodyne
detection over single photon counting at telecom wavelength has stimulated the study of continuous-variable protocols in the last few years \cite{ralph,grosshans:nature}. However, despite its
technological advantages, continuous-variable QKD (CVQKD) is still not
considered as a true alternative to discrete QKD, mostly because it
seems restricted only to short distances. The main reason for that
lies in the classical post-processing of the data shared by Alice and
Bob who need to construct a key from continuous random values, which
is a task far more complicated than its discrete counterpart.

In this letter, we introduce a specific CVQKD scheme, which exhibits two specific related advantages~: first, it allows to simplify significantly both the modulation scheme and the key extraction  task, and second, it makes possible to distill  secret keys over much longer distances.

Continuous-variable protocols have recently be shown to be unconditionally secure, that is, secure against arbitrary attacks \cite{renner-cirac}. In particular, collective attacks are asymptotically optimal, meaning that the theoretical secret key rate $K$ obtained using one-way (reverse) reconciliation is bounded 
below by~:
\begin{equation}
  K\geq I(x:y)-S(y:E) \equiv K_{\text{th}},
\end{equation}
where $x,y$ represent the classical data of Alice and Bob, and  $E$ is Eve's quantum state. 
Here $I(x:y)$ refers  to the
Shannon mutual information \cite{cover} between classical random
values $x$ and $y$, and $S(y:E)$ is the quantum mutual information
\cite{Nielsen} between $y$ and the quantum state $E$. The reason for
using two different measures of information is that Eve has no
restriction on her capabilities (other than the ones imposed by quantum
mechanics), while Alice and Bob must be able to extract a key with  current
technology. This secret key rate is valid for reverse reconciliation \cite{grosshans:nature}:
the final key is
extracted from Bob's data and Bob sends some side-information to Alice on
the authenticated classical channel to help her correct her errors.
In addition, one should note that $K_{\text{th}}$ 
corresponds to a scenario where Alice and Bob could perform perfect
error correction, which is never the case in practice. For this
reason, the key rate must be modified in the following way
\cite{lodewyck:pra07,leverrier:pra08}~:
\begin{equation}
  K_{\text{real}} = \beta \; I(x:y)-S(y:E)
\end{equation}
where $\beta$ is the so-called {\em reconciliation efficiency}. 
The term $\beta \; I(x:y)$ simply corresponds to
the amount of information Alice and Bob have been able to extract through
reconciliation. The second term, $S(y:E)$, is
bounded from the correlation between Alice and Bob's data, using an Heisenberg-type inequality.

Whereas the reconciliation efficiency is not usually taken into account to estimate asymptotic bounds, we 
must include it in our analysis because it is the currently limiting factor for the range of CVQKD with Gaussian modulation. In \cite{leverrier:pra08}, it was argued that working at low signal-to-noise ratio (SNR, defined as
the ratio of Alice's modulation variance to the noise variance) 
increases the range of the protocol. Unfortunately, maintaining a good reconciliation efficiency at very low SNR is 
even more difficult to achieve. This point is exactly the limitation that the protocol presented in this paper manages to overcome, hence allowing QKD over longer distances.

The paper is organized as follows~: after detailing the limitations of the Gaussian modulation, we present our new four-state protocol as well as its unconditional security proof. Then we describe the reconciliation step and show that its efficiency remains remarkably high, even at very low SNR. Finally, we show the expected performances of the protocol and  discuss some perspectives.

{\it Gaussian vs discrete modulation.}
Most CVQKD protocols use a Gaussian modulation since
it is the one maximizing the mutual information between Alice and
Bob over a Gaussian channel. In such a protocol, Alice draws two random values $q_A, p_A $ with
a Gaussian distribution $\mathcal{N}(0,V_A)$ and sends a coherent
state $|q_A + i p_A \rangle$ to Bob. The main problem of this
modulation arises when one wants to perform QKD over long
distances. In this case, there are two possibilities to fight the
noise induced by the losses in the channel~: either increase the
variance of the modulation so that the SNR
remains reasonably high, or work at low SNR. 
Unfortunately both approaches tend to fail over a few
tens of kilometers.

Working at high SNR requires to achieve a very good reconciliation efficiency, 
otherwise the secret key rate goes to zero \cite{lodewyck:pra07,leverrier:pra08}. Capacity-achieving error correcting codes are therefore required for this
task. Unfortunately, even with the best codes presently available
(LDPC codes \cite{Richardson2001a} or turbo codes \cite{berrou}),
one cannot expect to extend the range of the protocol well over 30
kilometers \cite{lodewyck:pra07}.

Working at low SNR relieves a little bit the need for capacity
achieving codes, but reasonably good low-rate codes are still hard to combine with the Gaussian modulation. Some interesting
algebraic properties of $\mathbb{R}^8$ can be useful in this
situation, and help increasing the achievable distance to over 50
kilometers \cite{leverrier:pra08}.

At present time both these approaches seem to have been pushed at
their maximum using the state-of-the-art channel coding techniques, and
breaking this 50 kilometers limit seems unlikely with a Gaussian
modulation.

One should now emphasize the following point~:
if the optimality of the Gaussian modulation over binary modulation is
clear at high SNR (simply because a binary modulation does not allow
to send more than 1 bit of information per signal), it is not true any
longer for low SNR. Adding to this fact
that a binary modulation allows for a much better reconciliation efficiency at low SNR, we infer that the modulation required to
achieve long distances is not Gaussian. Examples of binary
(or quaternary depending on the number of quadratures considered)
modulation have been proposed in the past
\cite{hirano2003qcu,namiki2006epe} but often combined with a
postselection procedure \cite{silberhorn2002}, and are not known to be unconditionally secure. 

{\it The four-state protocol.}
The protocol we propose runs as follows. Alice sends randomly one of the four
coherent states~: $|\alpha e^{i(2k+1)\pi/4} \rangle$ with $k \in \{0,1,2,3\}$. The amplitude $\alpha$ 
(taken as a real number) 
is chosen so as to
maximize the secret key rate one can expect from the expected
experimental parameters (transmission of the line and excess noise). Bob
measures randomly one of the quadratures in the case of the homodyne
protocol \cite{note} and gets the result $y$. 
The sign of $y$  encodes the
bit of the raw key while Bob reveals the absolute value $|y|$ to Alice
through the classical authenticated (but not secure) channel. At this point, Alice and Bob share correlated strings of bits. In order to help Alice correct her data, Bob sends some side-information over the classical channel, typically the
syndrome of his string relative to a binary code they agreed on
beforehand. From a classical communication perspective, the error correction (reconciliation) is then a problem of channel coding for the so-called {\em BIAWGN channel}, where a
binary modulation is sent over an Additive White Gaussian Noise channel, and for which there exist very good codes, even for extremely low
SNR.

The present protocol can thus be seen as an hybrid between the Gaussian modulation protocol,  with which it shares the physical implementation as well as the security proofs based on the optimality of Gaussian states, and protocols combining a discrete modulation with postselection, for which the error correction is substantially easier to perform, but whose unconditionnal security has not yet been established.

Let us now prove that the four-state protocol is unconditionally secure. First, it is enough to prove the security against collective attacks as they are the most powerful attacks in the asymptotic limit \cite{renner-cirac}. Then, as usual, the security is established by considering the equivalent entanglement-based version of the protocol. The state sent to Bob in the {\em prepare and measure} scheme is a mixture of four coherent states: $\rho =  \frac{1}{4} \sum_{k=0}^3 |\alpha_k\rangle\langle \alpha_k|$ with $\alpha_k = \alpha \exp(i(2k+1)\pi/4)$. 
The entanglement-based version uses a purification $|\Phi\rangle$ of
this state such that~: $\rho =  \text{tr}_A(|\Phi\rangle \langle \Phi|)$.
This state $\rho$ can be diagonalized as $$\rho = \lambda_0 |\phi_0\rangle \langle \phi_0|+\lambda_1 |\phi_1\rangle \langle \phi_1|+\lambda_2 |\phi_2\rangle \langle \phi_2|+\lambda_3 |\phi_3\rangle \langle \phi_3|$$
where  $\lambda_{0,2} = \frac{1}{2} e^{-\alpha^2}(\cosh(\alpha^2) \pm \cos(\alpha^2))$,
$\lambda_{1,3} = \frac{1}{2} e^{-\alpha^2}(\sinh(\alpha^2) \pm \sin(\alpha^2))$ and $$|\phi_k\rangle =\frac{e^{-\alpha^2/2}}{\sqrt{\lambda_k}}\sum_{n=0}^{\infty}\frac{\alpha^{4n+k}}{\sqrt{(4n+k)!}}(-1)^n|4n+k\rangle$$ for $k \in \{0,1,2,3\}$. 
Therefore, a particular purification of $\rho$ obtained by the Schmidt decomposition is $|\Phi\rangle = \sum_{k=0}^3 \sqrt{\lambda_
k} |\phi_k\rangle |\phi_k\rangle$
which can be rewritten as $|\Phi\rangle = \frac{1}{2} \sum_{k=0}^3 |\psi_k \rangle |\alpha_k \rangle$ where the states $$|\psi_k \rangle = \frac{1}{2} \sum_{m=0}^3  e^{-i(1+2k)m \frac{\pi}{4} } |\phi_m\rangle$$ are orthogonal non-Gaussian states.

The entanglement-based version of the four-state protocol can be described as follows. The state $|\Phi\rangle$ is distributed to Alice and Bob through a channel with transmission $T$ and excess noise $\xi$. Alice performs the projective measurement $\{|\psi_0\rangle\langle \psi_0 |,|\psi_1\rangle\langle \psi_1 |,|\psi_2\rangle\langle \psi_2 |,|\psi_3\rangle\langle \psi_3 | \}$ on her state, and thus prepares the coherent state $|\alpha_k \rangle$ when her measurement gives the result $k$.

\begin{figure}
  \includegraphics[width=.5\linewidth]{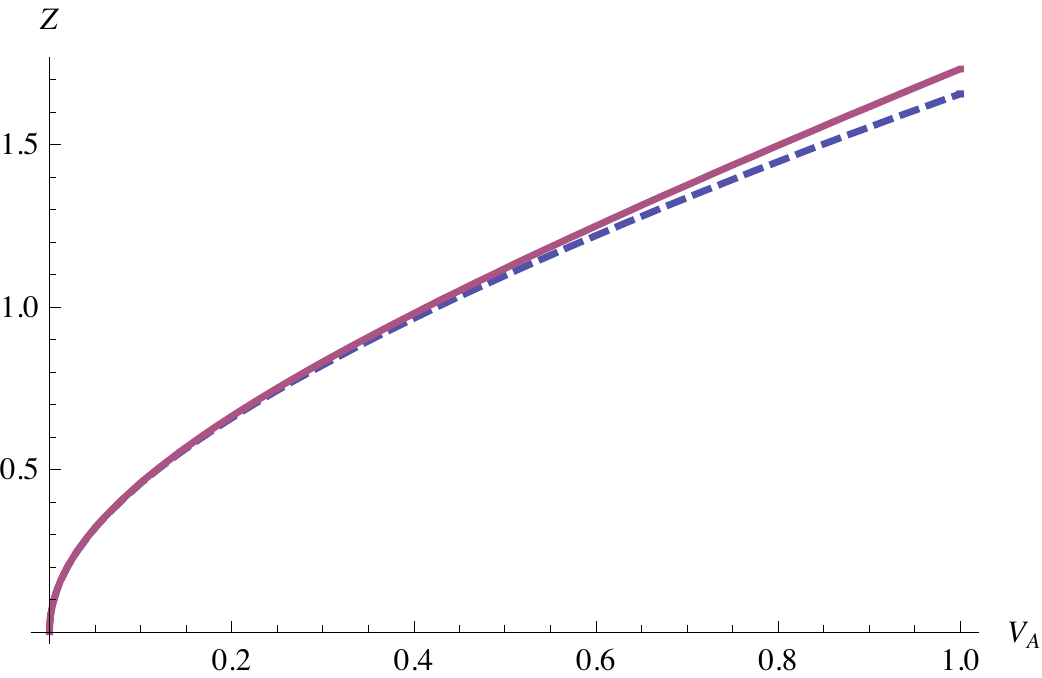}\hfill
  \includegraphics[width=.5\linewidth]{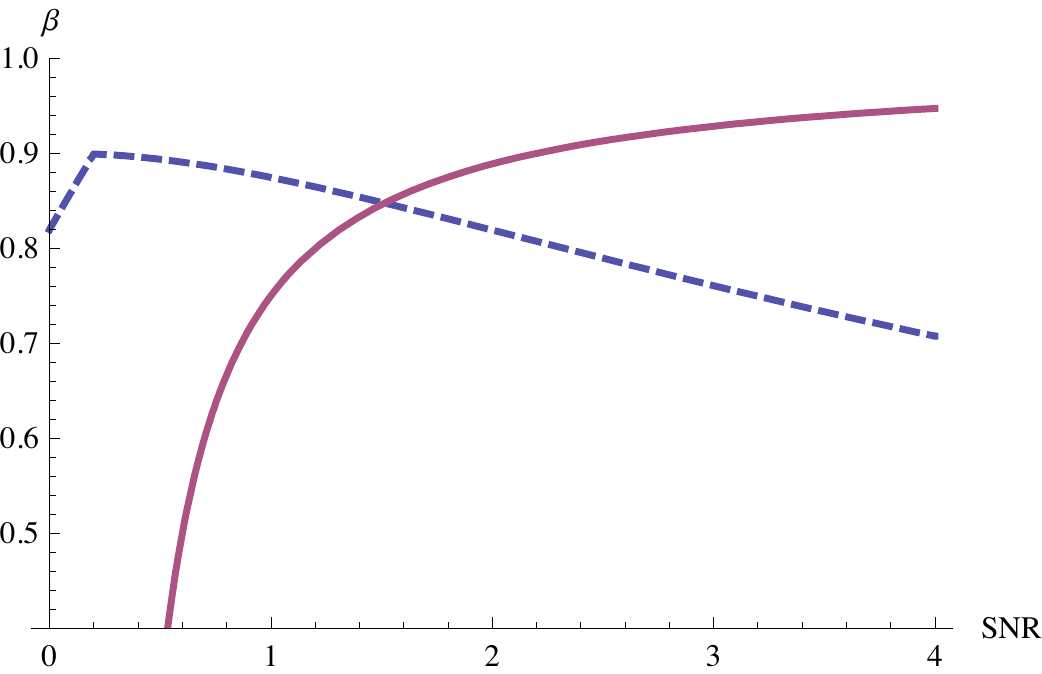}
  \caption{(Color online) Left~: correlation  $Z_{\text{EPR}}$ of an EPR pair (full line) and correlation $Z$  of state  $|\Phi\rangle$ (dashed line) as a function of the modulation variance.  Right~: reconciliation efficiency for a binary modulation (dashed line)  and for a Gaussian modulation (full line) \cite{Bloch2006}.}
  \label{comparison}
\end{figure}

In order to prove the security of this protocol, we use the extremality of Gaussian states to bound the Holevo information between Eve and Bob's classical variable \cite{garcia-patron:prl,navascues:prl06}. 
This Holevo information can be computed from the covariance matrix $\Gamma$ of the state shared by Alice and Bob~:
\begin{displaymath}
\Gamma = 
\begin{pmatrix}
(V_A+1) \openone_2  & \sqrt{T} Z \sigma_z\\
\sqrt{T} Z \sigma_z & (T V_A+1 + \xi) \openone_2\\
\end{pmatrix}
\end{displaymath}
where $V_A$ is the variance of Alice's modulation in the {\em prepare and measure} scheme and $T$ and $\xi$ refer to the experimentally estimated transmission and excess noise of the channel.
This covariance matrix has the same form as in the Gaussian modulation scheme where $Z$ would be replaced by  $Z_{\text{EPR}} = \sqrt{V_A^2+2V_A}$, the correlation of an EPR pair. The correlation $Z$ for the state $|\Phi \rangle$ does not take such a simple mathematical form but turns out to be almost equal to $Z_{\text{EPR}}$ for small variances (see Fig.~\ref{comparison}).

This means that if Alice uses a modulation with small variance, then all the information shared between Alice, Bob and Eve will essentially be the same as in the Gaussian modulation case, so the secret rate will also be the same \cite{lodewyck:pra07}. Combining this with an efficient reconciliation at low SNR (see Fig.~\ref{comparison}) allows to distill a key in conditions where the Gaussian modulation protocol is ineffective.

{\it Realistic reconciliation.} The main advantage of a binary modulation compared to a Gaussian
modulation is that one can find binary codes allowing high reconciliation efficiency, {\em e.g.} $80
\%$, even with a SNR close to 0. This is quite remarkable since the
reconciliation efficiency for a Gaussian modulation dramatically
drops to zero as the SNR becomes too low (see Fig.~\ref{comparison}).
In order to achieve an efficient reconciliation at low SNR, one needs
good low-rate codes. These can be constructed rather easily with a concatenation of a capacity-achieving code and a repetition code that we describe now.

At the end of
the quantum exchange, Alice and Bob share two correlated vectors
$x=(x_1,\cdots, x_N)$ (with $x_i=\pm \alpha/\sqrt{2}$) and
$y=(y_1,\cdots,y_N)$. We will use the concatenation of a capacity-achieving code $C$ of length $m$ and a repetition code of length $k$, 
assuming that $N=mk$.  Bob starts by defining the vector $Y=(Y_1,
\cdots, Y_m)$ where $Y_i=\text{sign}(y_{k(i-1)+1})$ for $i \in \{1,\cdots,m\}$. The goal of the reconciliation is for Alice to be able
to compute the vector $Y$. To do this, Bob sends some
side-information~: the vector $\{|y_1|,\cdots,|y_N|\}$, the $m$ vectors
$\{(1, \text{sign}(y_{k(i-1)+1} \times y_{k(i-1)+2}), \cdots, \text{sign}(y_{k(i-1)+1} \times y_{ki})\}$, and the syndrome of $\tilde{Y}$ for the code
$C$. This scheme allows Alice and Bob to extract $m$ bits out of their $N=km$ data.

This repetition scheme is a simple way to build a good code of rate $R/k$ out of a code of rate $R$. 
This construction is not optimal compared to
using a very good error correcting code at the considered
signal-to-noise ratio but exhibits some interesting features. First, designing very good codes at low
SNR is not easy, and has not been intensively studied so far, mainly
because the telecom industry does not operate in this regime~: this
would not be economical since an important number of physical signals
would be required to send one information bit. The problem is very
different in QKD,  where quantum noise is an advantage rather than a drawback. A second advantage of this repetition
scheme lies in its simplicity. As we mentionned earlier, the main
bottleneck of CVQKD is the reconciliation~: it was
limiting both the range and the rate of the
protocol. In particular, the rate is limited by the complexity of decoding LDPC codes, which  
is roughly proportional
to the size of the code considered (in fact $O(N \log N)$). If one uses a
repetition scheme of parameter $k$, then the length of the LDPC code
becomes $m=N/k$ allowing a speedup of a factor $k$.  The speed of the
reconciliation is not proportional to the number of signals exchanged
by Alice and Bob anymore, but to the mutual information they share,
which is a major improvement for noisy channels, {\em i.e.}, long
distance. Finally, the penalty in terms of reconciliation efficiency
imposed by using this scheme instead of a dedicated low rate error
correcting code is actually quite small. 
Roughly speaking, a repetition code of length $k$ allows to decode at a SNR $k$ times smaller. It is indeed easy to show that the efficiency $\beta_R(s/k)$ obtained at a SNR $s/k$ with such a repetition code is related to the efficiency $\beta_{LDPC}(s)$ available at SNR $s$ through $\beta_R(s/k)=\frac{\log_2(1+s)}{k
  \log_2(1+s/k)}\beta_{LDPC}(s)$, that is,  $\beta_R(s/k) \approx
(1-\frac{s}{2})\beta_{LDPC}(s)$ when $s$ is small enough. 
For instance,
there exist good LDPC codes of rate $1/10$ decoding at SNR $0.17$
\cite{richardson2002met}, meaning that $\beta_{LDPC}(0.17) \approx 88 \%$ and
$\forall k \geq 1$, $\beta_R(\frac{0.17}{k}) \geq 80 \%$. By using
different codes, one can have a reconciliation efficiency greater than
$80 \%$ for all SNR below 1.

The reconciliation scheme presented above performs indeed much better at low SNR (lower that 1) than reconciliation schemes used for a Gaussian modulation. This behaviour is inverted for higher SNR as a binary modulation is unable to send more that one bit per channel use. As a consequence, the four-state protocol is particularly relevant in a long distance scenario, whereas the Gaussian modulation protocol is still better suited to distribute high key rate at short distances.

\begin{figure}[th]

  \centerline{
    \includegraphics[width=8.6cm]{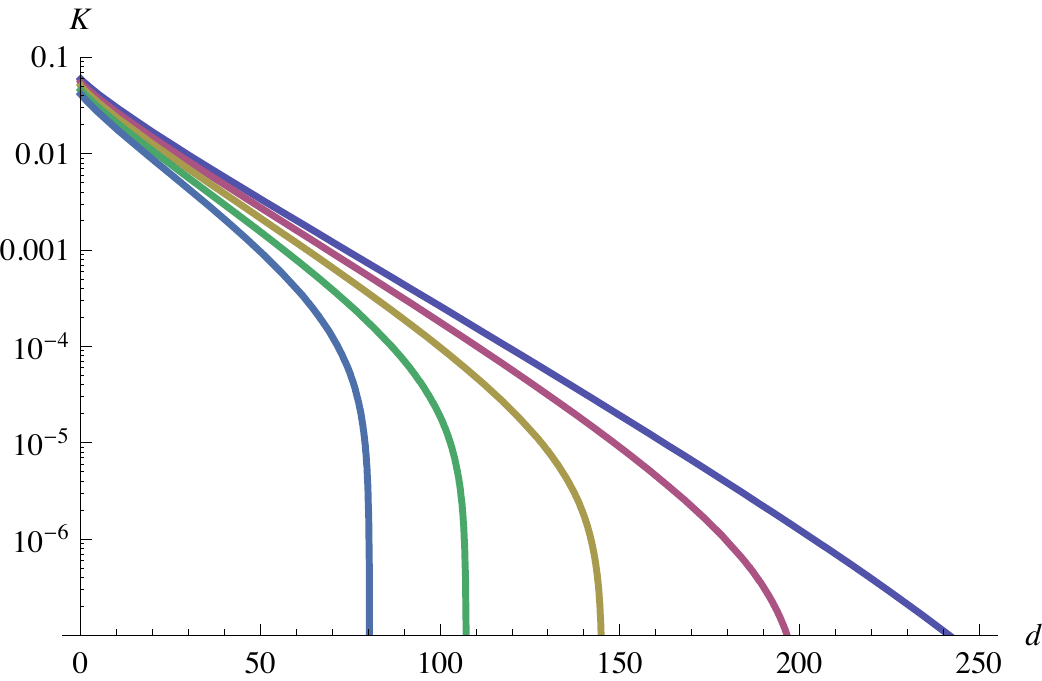}}
  \caption{(Color online) Secret key rate as a function of the distance for different values of the excess noise: from top to bottom, $\xi = 0.002, 0.004, 0.006, 0.008, 0.01$. The quantum efficiency of Bob's detection  is $\eta=0.6$.}
  \label{perfs}
\end{figure}

{\it Results and perspectives.} The theoretical performances of the four-state protocol are displayed on Fig.~\ref{perfs}. The quantum channel is characterized by its transmission $T = \eta  \; 10^{-0.02 d}$ where $\eta$ is the quantum efficiency of the homodyne detection and $d$ is the distance between Alice and Bob, and its excess noise $\xi$, that is the noise in excess compared to the shot noise. It should be noted that these performances are comparable with discrete-variable protocols, and are much better than previous CVQKD schemes.

Whereas Alice usually sends coherent states with a few photons per pulse (between 3 and 10) in the Gaussian modulation protocol, here, the optimal number of photons per pulse typically ranges from 0.2 to 1.
Therefore, the similitudes with discrete-variable QKD are important~: the information is encoded onto low amplitude coherent states with generally less than 1 photon per pulse. The main difference is that an homodyne detection replaces  photon counting. In our protocol, however, the error rate is not upper bounded (and can be as close as 0.5 as the reconciliation efficiency allows). This sounds in disagreement with security proofs for discrete-variable protocols that impose a maximum admissible quantum bit error rate (QBER). The reason for which this is nonetheless correct is that the error rate in our case in induced by both the noise added by Eve as well as the losses. This is in fact equivalent to a BB84 protocol where Bob would give a random value to each pulse he did not detect. In this case, the QBER is arbitrarily high,  but the security is still insured. 
In some sense, the main difference between the two schemes is that the vacuum noise is processed in two very different ways~: whereas it creates ``deletion errors'' (which are ignored) in the photon counting scheme, it produces ``real errors'' (which have to be corrected) in the continuous-variable scheme. But in both cases, these errors due to vacuum noise cannot be exploited by anybody, neither Alice or Bob, nor Eve.

As a conclusion, we presented a new unconditionally secure continuous-variable QKD protocol based on a discrete modulation. The use of good error correcting codes at low SNR allows to achieve long distances, which was impossible with a Gaussian modulation. Further work will include analysis of the finite-key effects \cite{scarani:prl08}, as well as the implementation of the present protocol.


\begin{acknowledgments}
We thank Joseph Boutros, Nicolas Cerf and Norbert L\"utkenhaus  for helpful discussions. We  acknowledge
  support from the  European Union under project SECOQC
  (IST-2002-506813), and from Agence Nationale de la Recherche
  under projects PROSPIQ (ANR-06-NANO-041-05) and SEQURE
  (ANR-07-SESU-011-01).
\end{acknowledgments}


\begin{thebibliography}{18}
\bibitem{note} Alternatively, Bob can measure simultaneously both
quadratures with an heterodyne detection, 
and the security proof will proceed along the same lines.
\end{thebibliography}
\end{document}